# Reconstructing galaxy histories from globular clusters

**Michael J. West [1], Patrick Côté [2], Ronald O. Marzke [3] & Andrés Jordán [2]**

[1]Department of Physics & Astronomy, University of Hawaii, 200 W. Kawili Street, Hilo, Hawaii 96720, USA

[2]Department of Physics & Astronomy, Rutgers, The State University of New Jersey, 136 Frelinghuysen Road, Piscataway, New Jersey 08854, USA

[3]Department of Physics & Astronomy, San Francisco State University, 1600 Holloway Avenue, San Francisco, California 94132, USA

**Nearly a century after the true nature of galaxies as distant "island universes" was established, their origin and evolution remain great unsolved problems of modern astrophysics. One of the most promising ways to investigate galaxy formation is to study the ubiquitous globular star clusters that surround most galaxies. Recent advances in our understanding of the globular cluster systems of the Milky Way and other galaxies point to a complex picture of galaxy genesis driven by cannibalism, collisions, bursts of star formation and other tumultuous events.**

One way to study the history of galaxies is by observing those that are far away. Because light travels at a finite speed, we see distant objects as they looked in the past. By observing galaxies over a wide range of distances, astronomers are able to look back in time to see how galaxies were at different epochs, with the goal of someday constructing a complete chronology of galaxy evolution. In practice, however, the same great distances that make remote galaxies interesting targets for study also make them difficult to observe[1-3]. Additionally, the inherently statistical nature of this approach, although providing information about the evolution of galaxies as a population over cosmological timescales, can say little about the unique history of any particular galaxy[4-6].

An alternative way to learn about galaxy origins is to study those galaxies that are nearby. Because the properties of galaxies today reflect both the conditions at the time of their formation and the cumulative effects of billions of years of evolution, careful analysis of present-day galaxies can yield clues to their past. But the success of this approach hinges on the ability to identify some property of galaxies (or some 'tracer' population within them) that can serve as a reliable gauge of their evolution over billions of years. Here we focus on one of the most promising of tracer populations -- globular clusters. Remarkable progress has occurred over the past decade in our understanding of the globular cluster systems of galaxies, in particular the discovery that many galaxies possess two or more distinct subpopulations of globulars. An underlying theme of this Review is that astronomers are on the verge of reconstructing the individual formation histories of large numbers of galaxies from careful analysis of their globular cluster populations.

**Diagnostics of galaxy formation**

Globular clusters are dense, gravitationally bound collections of hundreds of thousands to millions of stars that share a common age and chemical composition. Most galaxies are surrounded by systems of tens, hundreds or even thousands of globular clusters that swarm about them like bees around a hive. An example is shown in Fig. 1. In general, the number of globular clusters that a galaxy possesses is roughly proportional to its luminosity[7].

The earliest studies of globular clusters were limited to those belonging to the Milky Way and its closest galactic neighbours, as the globular cluster systems of more distant galaxies were, with a few notable exceptions[8-10], too faint to be seen in significant numbers before the advent of sensitive electronic detectors for astronomical telescopes.

The ages of the Milky Way's approximately 150 globulars, which can be determined by comparing the s and luminosities of their constituent stars with models of stellar evolution, are found to be in the range 10 to 15 billion years[11]. This makes them among the most ancient objects in the Universe, fossils from the birth of our Galaxy that have survived to the present. For this reason, globular clusters have played a key role in the continuing debate over how the Milky Way formed. Searle & Zinn's analysis[12] of the globular cluster population in the halo of the Milky Way, for example, led them to conclude that the halo of our Galaxy formed via the slow accretion of many small proto-galactic fragments, contrary to the then-prevailing view of a more rapid and monolithic collapse[13]. Mounting evidence supports this view of the Milky Way as a cannibal that may have consumed dozens or perhaps hundreds of smaller galaxies or galactic fragments during its lifetime[14-18]. Such cannibalism appears to be quite common among galaxies[19,20]. Although this process eventually erases the structural identities of progenitor galaxies, their stars and globular clusters remain intact, providing a rich source of historical detail about the birth of present-day galaxies.

Globular clusters are especially attractive probes of galaxy formation for a number of reasons. Foremost is the evidence that some properties of globular cluster systems are correlated with those of their host galaxies[7,21,22], which suggests that the birth of these star clusters is linked to the formation of the galaxy itself and hence that they can be used to tease out information about the process of galaxy building. Furthermore, unlike the diffuse stellar population in galaxies, which spans a wide range of ages and chemical abundances, the stars within a given globular cluster are a coeval and extremely homogeneous population, which makes them much simpler stellar systems to understand. Additionally, their luminosity and compact size make them among the brightest and most readily identifiable objects within galaxies, allowing them to be traced to large galactocentric distances where observations of the diffuse stellar component of the galaxy become impractical because of the precipitous drop in surface brightness.

### Globular cluster subpopulations

Advances in astronomical imaging and spectroscopic capabilities over the past decade, especially the exquisite image quality provided by the Hubble Space Telescope, have led to tremendous progress in our understanding of the globular cluster systems of galaxies beyond the Milky Way. Two key discoveries in particular have forced a rethinking of the earlier view of globular clusters as a homogeneous population of ancient fossils leftover from the earliest stages of galaxy formation.

First, the detection of young massive star clusters in starburst and merging galaxies whose sizes, luminosities and colours suggest that they may be newly formed globular clusters[23-27]. The presence of such objects in disturbed galaxies suggests that the disturbances themselves may act as a catalyst for the birth of globular clusters, and raises the question of whether this might be the primary mechanism of cluster formation. Age dating of such clusters can provide constraints on the frequency and duration of galaxy interactions[28,29].

This first discovery was soon followed by the second: namely, that many – perhaps most – large galaxies possess two or more subpopulations of globular clusters that have quite different chemical compositions[30-33]. Figure 2 shows histograms of globular cluster metallicities for several nearby elliptical galaxies. By convention, the metallicity of stars and star clusters, which astronomers denote [Fe/H], is measured relative to the Sun's chemical abundance

$$\left[\frac{\text{Fe}}{\text{H}}\right] = \log\left(\frac{N_{Fe}}{N_H}\right) - \log\left(\frac{N_{Fe}}{N_H}\right)_{Sun}$$

where $N_{Fe}/N_H$ is the relative numbers of iron and hydrogen atoms, a good indicator of overall metallicity. Globular cluster metallicities are usually inferred from their photometric colours, which serve as a proxy for the more reliable but also more observationally challenging spectroscopic measurement of chemical abundances[34-39]. Two peaks are clearly seen in the globular cluster metallicity distribution of each galaxy in Fig. 2, and statistical modeling confirms that these are well described by the sum of two gaussian distributions with distinct metal-poor and metal-rich peaks. Subsequent studies revealed that these different globular cluster subpopulations also have different spatial distributions and kinematics[40-45].

With hindsight, the ubiquity of multiple globular cluster populations should perhaps not have come as any great surprise, as two distinct globular cluster subsystems – both old – have long been known to co-exist within the Milky Way[46-49]. The majority of Galactic globular clusters belong to a metal-poor ([Fe/H] < -0.8) population that resides in a slowly rotating, spherical halo surrounding our Galaxy. A second, metal-rich ([Fe/H] > -0.8) population shares the flattened spatial distribution and rapid rotation characteristic of the bulge and disk components of the Milky Way.

### Competing formation scenarios

The presence of multiple globular cluster subpopulations within a single galaxy is undoubtedly an important clue to how such galaxies formed. A number of different hypotheses have been proposed to explain this phenomenon.

One possibility is that the metal-poor and metal-rich populations correspond to different generations of globular clusters. In the simplest cosmological theories of structure formation, galaxies are envisioned as having originated billions of years ago from clouds of metal-poor gas, mostly hydrogen and helium, that collapsed under gravity's pull to form stars and clusters of stars, the most massive of which might today be identified as globular clusters. Over time, supernovae -- the explosive deaths of massive stars – enrich any remaining intragalactic gas with the thermonuclear 'ashes' from their interiors, such as carbon, oxygen, and other heavier atoms. If a second generation of globular clusters forms from this enriched gas, then two separate species of clusters would co-exist within the same host galaxy, one younger and more metal rich than the other.

But how to trigger multiple episodes of globular cluster formation? Mergers of gas-rich galaxies is one possible mechanism[50-53], as shock waves generated by the collision might compress the gas and lead to the birth of new globular clusters to join any extant population. This idea received strong support with the discovery of newly formed massive star clusters in some interacting galaxies that might be young globular clusters . An example is shown in Fig. 3. Even without collisions as a trigger, it has been suggested that the *in situ* formation of multiple generations of globular clusters might occur under certain conditions within a single galaxy[54-56].

A fundamentally different hypothesis for the origin of globular cluster subpopulations in galaxies is to assume that all clusters are roughly coeval and old, and that their numbers are conserved during the hierarchical assembly of galaxies. If, as evidence suggests, large galaxies have grown to their present sizes by cannibalizing smaller ones (see, for example, Fig. 4), or by accreting material stripped from other galaxies during collisions, then the resulting globular cluster systems of such galaxies will be an admixture of their original populations plus those inherited from their victims[57-61]. Given the preponderance of small galaxies in the universe, and the observed correlation between galaxy mass and the mean metallicity of its globular clusters[21,59] most of the inherited globulars would be more metal-poor than those of the original progenitor galaxy. Galaxy cannibalism and accretion should thus lead to chemically distinct subpopulations of old globular clusters in most large galaxies. Numerical simulations confirm that multi-peaked metallicity distributions like those observed arise quite naturally in such a scenario without requiring the formation of multiple generations of clusters[59-61]. Mounting evidence of tidal streams of stars and gas[15-20,62,63] in the Milky Way and other nearby galaxies -- the vestiges of cannibalized companions – lends support to the notion that a significant fraction of their present-day globular cluster populations may have once resided in other galaxies.

**Model strengths and weaknesses**

Although each of the competing models offers a plausible explanation for the existence of multiple globular cluster populations within the same host galaxy, each also suffers from a number of shortcomings.

Models that assume the creation of multiple generations of globular clusters as a result of galaxy collisions or other episodic bursts of star formation are predicated on processes whose physics is poorly understood at present, such as gas heating and cooling, globular cluster formation, the effects of feedback from star formation, and the complex interplay between them. Producing two or more globular cluster subsystems with markedly different metallicities within the same galaxy also requires some fine-tuning. If, for example, globular cluster formation is a prolonged process rather than occurring in bursts at two or more well separated epochs in the metal enrichment histories of galaxies, then the resulting cluster populations would likely have a broad distribution of metallicities lacking the distinct peaks seen in the majority of large galaxies today. Such models are also strongly constrained by the very old ages found for both the metal-rich and metal-poor cluster populations in several of the most thoroughly studied galaxies[35,39,64-67]. The apparent paucity of young and intermediate aged clusters suggests either that the bulk of galaxy mergers and major star formation episodes must have been completed many billions of years ago, or that such events are not the primary mechanism by which most clusters form.

The accretion model of pre-formed globular clusters also faces a number of challenges. Most obviously, it ignores the creation of new globular clusters that is clearly taking place in some interacting galaxies today, relegating this to a second-order effect rather than the primary formation mechanism for most clusters. Additionally, the accretion model requires a distribution of proto-galactic masses that was very heavily weighted in favour of low-mass objects in order to produce globular cluster metallicity distributions similar to those seen in Fig. 2, with large galaxies having been assembled from hundreds or thousands of cannibalized smaller ones[59-61]. Such a steep mass function is inconsistent with observations of the luminosity function of galaxies today[68-71], but is suggestively similar to the primordial mass spectrum predicted by the popular cold-dark-matter models of structure formation.

There is, of course, the possibility that the present-day globular cluster populations of galaxies might have arisen from some combination of collisions, cannibalism and accretion, or perhaps from other processes. For example, a number of authors have suggested that the stripped nuclei of compact dwarf galaxies might masquerade as globular clusters[72-74]. A good nearby example is the Milky Way's largest globular cluster, ω Centauri, which many astronomers now suspect may be the surviving nucleus of a dwarf galaxy that was gravitationally shredded by our galaxy[75,76]. Additional complications come from recent discoveries of a new class of faint 'fuzzy' star clusters[77], 'ultra compact dwarf' galaxies[78] and a population of intergalactic globular clusters of unknown origin[79,80]. Although these various objects probably comprise only a small fraction of the globular clusters in normal galaxies, their existence demonstrates that the processes leading to the formation of present-day globular cluster systems are likely to have been many and varied.  The key question is what was the dominant process that governed the assembly of present-day globular cluster systems.  One conclusion shared by the various competing models is that the genesis of present-day galaxies and their globular cluster systems has been a long, drawn-out process that is still ongoing, rather than a single event in the distant past.

**Where do we go from here?**

Much work remains to be done to better understand extragalactic globular cluster systems and their use as probes of galaxy formation.  Significant progress in the near future is likely to occur along several fronts.

On the theoretical front, each of the competing models for the origin of multiple globular cluster populations in galaxies is undoubtedly oversimplified.  New insights may come from high-resolution computer simulations or semi-analytic models of galaxy formation that start from more realistic cosmological initial conditions. By following the detailed merger histories of simulated galaxies from early epochs to the present, and including the effects of gas dynamics, star formation, chemical enrichment, and energy input from supernovae, it should be possible to refine quantitative predictions about the evolution of globular cluster systems as a function of time.  Important first steps in that direction have already been taken[81-86], and within the next few years it is hoped that numerical simulations will achieve sufficient mass resolution to allow the growth of globular clusters, galaxies and larger-scale structures to be followed simultaneously.

On the observational side, although the globular cluster systems of distant galaxies will likely remain inaccessible to direct observation for the foreseeable future[87], many thousands of nearby galaxies within the grasp of current technology await study.  Thanks to the new generation of large, sensitive CCD cameras, including the new *Advanced Camera for Surveys* onboard the Hubble Space Telescope, it is feasible to obtain deep, multi- images of large numbers of galaxies with a modest investment of telescope time.  A comprehensive census of the globular cluster systems of galaxies in the local universe would be invaluable, as it would provide a more thorough assessment of the frequency of globular cluster subpopulations, their properties, and variations with galaxy type and environment.  An example of what can be done is the recent *ACS Virgo Cluster Survey* (http://physics.rutgers.edu/~pcote/acs), which will provide data for many thousands of globular clusters associated with 100 early-type galaxies in the nearby Virgo Cluster of galaxies.   Galaxies that do not have multiple globular cluster populations might also provide insights to the galaxy formation process.  Direct comparison of the metallicities of globular clusters and halo stars in the same galaxy, although observationally challenging, is likely to shed additional light on the origin of galaxies[88,89].

Major spectroscopic programmemes are also needed to obtain large numbers of radial velocities and abundance measurements of globular clusters from high signal-to-noise spectra; such data would allow more detailed studies to be undertaken of the kinematics, compositions and star formation histories of cluster subpopulations.  Although photometric observations can be, and commonly are, used to infer the metallicities of globular clusters, they are no substitute for spectroscopy.  Spectroscopic measures of metallicity are much more precise than the photometric measures described earlier, and our ability to produce precise timelines is tied directly to the accuracy of the metallicities.  However, the faintness of all but the nearest clusters makes such observations challenging even with telescopes of the 8-10 meter class. The next generation of very large 30 to 50 m telescopes that are planned for the coming decade should allow great progress to be made in this area.

An especially fertile area of research in the next few years may be age determinations of globular cluster subpopulations in galaxies, as here the competing models make rather different predictions.  If the metal-rich and metal-poor clusters are the result of two separate bursts of cluster formation, then the metal-rich globulars must be younger than those that are metal-poor.  If, on the other hand, the multiple globular cluster populations seen in galaxies can be accounted for by the accretion of pre-existing globulars, then the metal-rich and metal-poor globulars should be predominantly old and coeval, with no obvious correlation between metallicity and age.  Whereas current methods of age determinations are not sufficiently precise to rule out the competing models[35,39,64-67], improved methods hold great promise for reducing uncertainties to the point where stronger constraints on models can be made.  Detailed spectroscopic and photometric observations of the most massive galaxies in dense environments show that their globular cluster populations are generally old and coeval to within a couple of billion years[35,39,64-67].  But the age distribution of globular clusters remains an open question, as several recent studies have found a wider spread of globulars cluster ages in some galaxies, including the presence of some intermediate-age metal-rich clusters[90-92]. The findings to date suggest a complex picture of globular cluster formation in galaxies, with the formation histories of no two galaxies being exactly alike.

These are exciting times in the study of extragalactic globular cluster systems.  Disentangling the myriad processes that may have contributed to the formation of nearby galaxies is a formidable challenge, but the potential reward is spectacular: a glimpse into the evolutionary histories of individual galaxies.

**Acknowledgements:** We thank J. Brodie, W. Harris and S. van den Bergh for comments that helped to improve this manuscript.  Correspondence and requests for materials should be addressed to M.W. (westm@hawaii.edu).

**FIGURES**

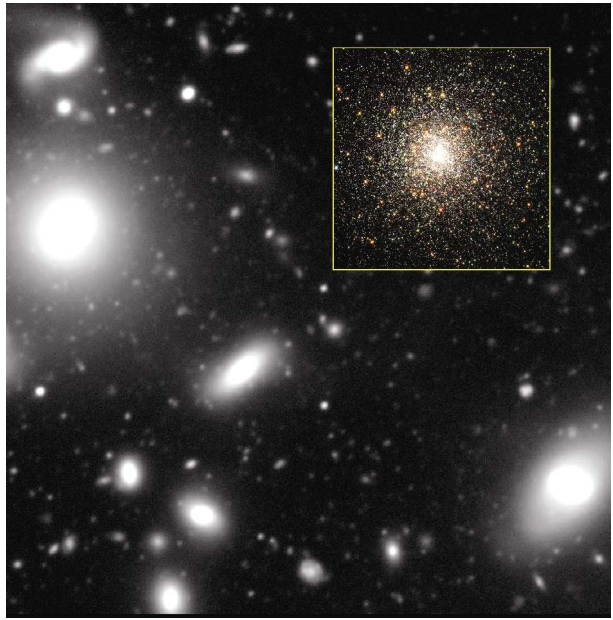

Figure 1. Extragalactic globular cluster systems. Globular clusters are seen as a swarm of faint point-like sources surrounding galaxies in this deep image obtained with the Keck I telescope. Inset, a Hubble Space Telescope image of a nearby globular cluster in our own Milky Way galaxy. A typical globular cluster is composed of as many as a million stars. Most galaxies possess globular clusters systems, from dwarf galaxies with just a handful of clusters to giant galaxies with tens of thousands of clusters. Inset image courtesy of NASA and the Hubble Heritage Team (STScI/AURA).

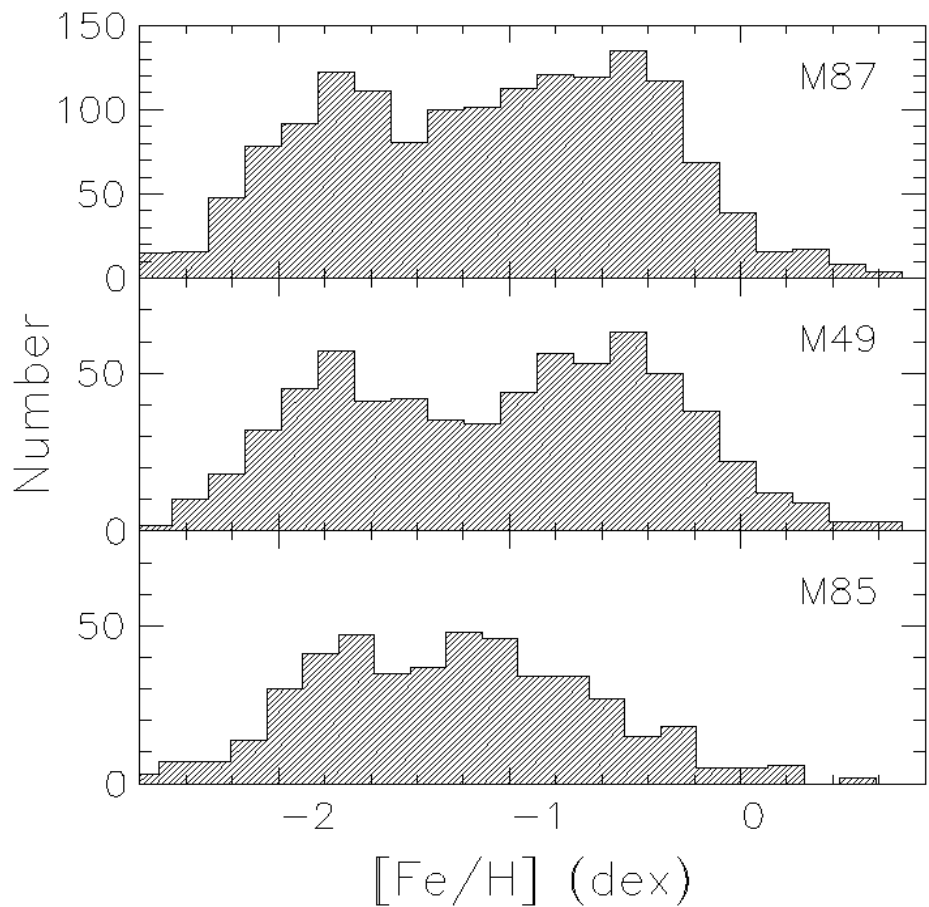

Figure 2. Histograms of globular cluster metallicities in three galaxies. The metallicity, [Fe/H], has been inferred from photometric colours of each galaxy's globular clusters obtained by the authors using the Hubble Space Telescope. Two or more distinct peaks are seen in each case, which indicates the presence of multiple populations of globular clusters with different chemical compositions within these galaxies. A number of different theories have been proposed to explain the origin of these multi-modal globular cluster distributions (see text).

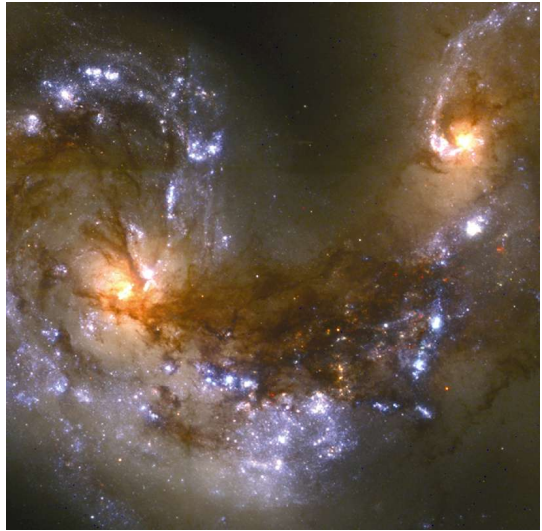

Figure 3: Hubble Space Telescope image of a colliding pair of galaxies. The collision appears to have sparked a recent burst of star formation, including the birth of over 1,000 massive star clusters (the prominent blue clumps). The blue colour of these star clusters indicates ages of only a few million years. Some of the most massive objects may evolve over the next few billion years to resemble old globular clusters. This suggests that at least some globular clusters have been born as a result of galaxy collisions and mergers. Image courtesy of NASA/STScI/B.Whitmore.

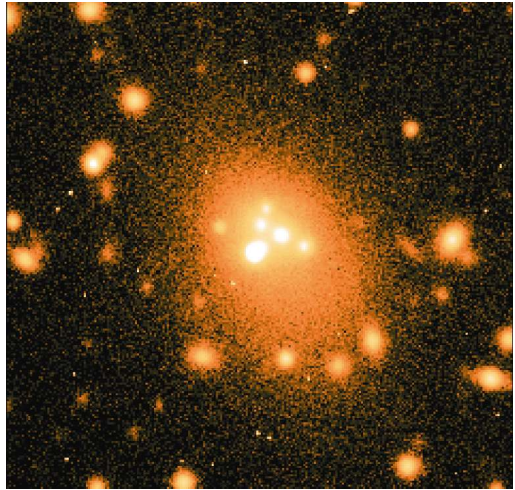

Figure 4: A giant elliptical galaxy in the cluster Abell 3827. The partially digested remains of several smaller cannibalized galaxies are visible in its central region. Globular clusters belonging to those galaxies are likely to survive the eventual disruption of their parent galaxies, and thus will become part of this giant galaxy. If, as evidence suggests, many large elliptical galaxies have grown to their present sizes by cannibalizing smaller neighbours, then their globular cluster populations today are composite systems that may provide clues about the progenitor galaxies.